\documentclass[twocolumn,aps,showpacs,showkeys,amsmath,amssymb,%
eqsecnum,preprintnumbers]{revtex4} 

\usepackage{dcolumn}
\usepackage{bm}
\def\Tr{\mbox{Tr}\,}
\def\W{\wedge}
\def\dag{^\dagger}
\def\D{\mbox{d}}
\def\half{\frac{1}{2}}
\def\dl{\delta\lambda}
\def\slash#1{\, /\kern-0.6em{#1}}

\begin{document}


\preprint{{hep-th/0109220}}


\title{Local symmetries of the non-Abelian two-form}     
\author{Amitabha Lahiri}
\email{amitabha@boson.bose.res.in}
\affiliation{S. N. Bose National Centre for Basic Sciences, \\
Block JD, Sector III, Salt Lake, Calcutta 700 098, INDIA}
\date{\today}

\begin{abstract}
It is proposed that a non-Abelian adjoint two-form in $B\W F$ type
theories transform inhomogeneously under the gauge group. The
resulting restrictions on invariant actions are discussed. The
auxiliary one-form which is required for maintaining vector gauge
symmetry transforms like a second gauge field, and hence cannot be
fully absorbed in the two-form. But it can be replaced, via a
vector gauge transformation, by the usual gauge field, leading to
gauge equivalences between different types of theories. A new type
of symmetry also appears, one which depends on local functions but
cannot be generated by constraints. It is connected to the identity
in the limit of a vanishing global parameter, so it should be
called a semiglobal symmetry. The corresponding conserved currents
and BRST charges are parametrized by the space of flat connections.
\end{abstract}

\pacs{11.30.Ly, 03.50.Kk, 11.10.Lm}

\keywords{Topological field theory, non-Abelian two form,
non-local symmetry\, }

\maketitle
\medskip

\section{\label{intro}Introduction}

The non-Abelian two-form, or antisymmetric tensor potential, first
made its appearance in the context of nonlinear
$\sigma$-models~\cite{Freedman:1977pa,Seo:1979id,Freedman:1981us}.
It was introduced as a Lagrange multiplier field via an interaction
term $\Tr \epsilon^{\mu\nu\rho\lambda} B_{\mu\nu}F_{\rho\lambda}$,
usually written more compactly in the notation of forms as
$\Tr\,B\W F\,.$ Here $B$ is a two-form potential in the adjoint
representation of the gauge group, and $F$ is the curvature of the
gauge connection, i.e., the field strength of the gauge field. An
action made up of this term alone is a Schwarz type topological
field theory~\cite{Schwarz:1978cn,Schwarz:1979ae,Witten:1988ze,
Birmingham:1991ty}. It generalizes to four dimensions the
Chern-Simons action, a well known topological quantum field theory
in three
dimensions~\cite{Schonfeld:1981kb,Deser:1982vy,Deser:1982wh}.

This action serves as the cornerstone for a wide variety of
theories in four dimensions. It is diffeomorphism invariant, and
can be thought of as `topological gravity', a toy model for some
features of quantum
gravity~\cite{Horowitz:1989ng,Blau:1991bq,Baez:1994zz}. Some
approaches to quantum gravity use the fact that Einstein's theory
in four dimensions can be written in terms of the $B\W F$ action
with additional terms~\cite{Peldan:1994hi}, usually with
non-compact gauge groups. When the gauge group is taken to be
SU(N), modifications of the $B\W F$ action lead to Yang-Mills
theory in a first order
formulation~\cite{Fucito:1997ax,Cattaneo:1998eh, Fucito:1997sq,
Accardi:1997bf}, or in a loop space
formulation~\cite{Chan:1995bp,Chan:1996xr}. In these models the
two-form appears as a field without its own dynamics.

A theory where the two-form is dynamical can be constructed by
introducing a kinetic term~\cite{Lahiri:1992yz,Lahiri:1992hz}. This
also happens to be a non-Abelian generalization of a mass
generation mechanism for vector fields in four dimensions, which
does not have a residual Higgs particle in the
spectrum~\cite{Cremmer:1974mg,Aurilia:1981xg,Govindarajan:1982jp,
Allen:1990kc,Allen:1991gb,Minahan:1989vc}. The non-Abelian theory
can be quantized in the path integral approach using standard
algebraic
techniques~\cite{Hwang:1997er,Lahiri:1997dm,Lahiri:2001uc}. This
gives a formally renormalizable, gauge invariant theory which
contains massive vector bosons but does not have a residual Higgs
particle. Note that this mechanism evades established no-go
theorems because all those theorems restrict themselves, at some
point of the proof, to scalars and vectors only. But rewriting the
two-form purely in terms of those degrees of freedom would
inevitably lead to a nonlocal field theory, to which those proofs
do not apply.

Despite its wide applicability, the nature of the non-Abelian
two-form remains obscure. Much of the analysis of theories of the
two-form is based on quantization of its interactions. For example,
a two-form couples naturally to a world surface, so one possible
description of it is as a gauge field for strings, open or
closed. This is a consistent description for the Abelian
two-form. However, Teitelboim has shown~\cite{Teitelboim:1986ya}
that it is not possible to define `surface ordered' exponentials
for the non-Abelian two-form in a reparametrization invariant
fashion. This rules out a non-Abelian generalization of the picture
of the two-form as a connection for strings. In another approach,
consistency of deformations of theories with the $B\W F$
interaction suggests that it is not possible to construct theories
with terms quadratic in the non-Abelian two-form, unless additional
fields are introduced~\cite{Henneaux:1997mf}.

In this paper I investigate theories of the non-Abelian two-form
from another perspective, namely, that of classical internal
symmetries in the Lagrangian formalism. I will display several
classical symmetries of theories involving the two-form. These shed
some light on the geometrical nature of this field, although
precisely what that nature is remains unclear. One outcome of this
investigation will be to show that theories of the non-Abelian
two-form can be written as theories with two connections, one of
which is usual gauge field, and the other has the characteristics
of a {\em dual} connection. I will also show that only some of
these symmetries are generated by local constraints, a novelty for
classical field theories.

What sort of symmetry is imposed on the non-Abelian two-form will
dictate the structure of the resulting theory, with its own
physical interpretation. This will be the underlying theme of the
constructions in this paper, namely, to construct new symmetries by
extending known ones, and then to construct actions invariant under
them. Some results of this paper have been reported in brief
in~\cite{Lahiri:2001di}, many details and several new results are
presented in this paper.

The starting point is the interaction Lagrangian $ \Tr B\W F$. The
integral of this term will be called the action, somewhat loosely,
for all theories to be discussed in this paper are built around
this term. The symmetries of those theories will necessarily be
symmetries of this term alone until Section~\ref{twocon}, where
this term itself will need to be modified. The interaction
Lagrangian is often called a `topological' term, because a
background metric is not necessary to define it. However, the
metric independence of this term will not be crucial to the results
in this paper.  Only internal symmetries will be considered below,
and invariance of the action will mean invariance up to a total
divergence. 

In section~\ref{gauge} the non-Abelian two-form is shown to have an
inhomogeneous transformation law under the gauge group. Actions
invariant under these transformations are constructed in
section~\ref{actions}. Vector gauge transformations are introduced
in section~\ref{aux}, along with the auxiliary vector field, which
is shown to transform as a gauge field as well. The new semiglobal
transformations are also discussed there. The connection like
nature of the auxiliary field allows a fresh set of transformation
rules and actions, shown in section~\ref{twocon}, but they are
shown to be equivalent to the earlier ones. The main results are
summarized in section~\ref{last}.

{\em Notation:} It will prove convenient to use the notation of
differential forms. The gauge connection one-form (gauge field) is
defined in terms of its components as $A = -ig A^a_\mu t^a \D
x^\mu\,,$ where $t^a$ are the (hermitian) generators of the gauge
group satisfying $[t^a, t^b] = if^{abc}t^c$ and $g$ is the gauge
coupling constant. Any other coupling constant, which may be
required in a given model, will be assumed to have been absorbed in
the corresponding field. An example is a coupling constant $m$ of
mass dimension one, which appears in the action of some models as
$mB\W F$, and may be absorbed into $B$.  This will cause no problem
since I am dealing solely with classical systems and classical
symmetries, and I will display coupling constants explicitly
whenever the discussion requires it. The gauge group will be taken
to be SU(N). The gauge-covariant exterior derivative of an adjoint
$p$-form $\xi_p$ will be written as
\begin{eqnarray}
\D_A\,\xi_p \equiv \D\,\xi_p + A\W \xi_p + (-1)^{p+1} \xi_p\W A\,,
\label{0701.extderiv}
\end{eqnarray}
where $\D$ stands for the usual exterior derivative. The field
strength is $F = \D A + A\W A$, and satisfies Bianchi identity, $\D
F + A\W F - F\W A = 0$.  Under a gauge transformation, the gauge
field transforms as $A \to A' = UAU\dag +\, \phi$, where for later
convenience I have defined $\phi \equiv - \D UU\dag$. Note that
$\phi$ is a flat connection, $\D\phi + \phi\W\phi = 0$. Unless
otherwise noted, a transformed field will be distinguished by a
prime, for all types of transformations.

\section{\label{gauge}Gauge symmetries}
In this section I shall discuss the construction of a modification
of SU(N) gauge transformation rules for the non-Abelian two-form,
starting from the usual transformations.  As mentioned earlier, the
investigation of symmetries of the non-Abelian two-form $B$ has to
begin from the action $\int \Tr B\W F$. In terms of its components,
$B = -\half igm B^a_{\mu\nu}t^a \D x^\mu\W \D x^\nu$, with $m$ a
constant of mass dimension one. Under an SU(N) gauge transformation
represented by $U$, the gauge field $A$ and the field strength $F$
transform as
\begin{eqnarray}
A \to A' &=& UAU\dag + \phi\,,\nonumber \\
F \to F' &=& UFU\dag\,.
\label{0701.gauge1}
\end{eqnarray}
Invariance of the action under SU(N) gauge transformation is usually
enforced by assuming that $B$ transforms homogeneously in the
adjoint, as
\begin{eqnarray}
B \to B' = UBU\dag\,.
\label{0701.ugauge}
\end{eqnarray}
It should be noted that there is no {\em a priori} reason (based on
a geometrical description of the two-form, for example) to assume
this transformation law for $B$.  

In addition to SU(N) gauge transformations, the action is invariant
under a non-Abelian generalization of Kalb-Ramond gauge
transformation~\cite{Kalb:1974yc},
\begin{eqnarray}
B' = B + \D_A\xi, \qquad A' = A\,,
\label{0701.vector}
\end{eqnarray}
where $\xi$ is an arbitrary one-form. Because of Bianchi identity,
the Lagrangian changes by a total divergence under this
transformation. This will be referred to as vector gauge
transformation, while a transformation with $U$ will be called
(SU(N), or usual) gauge transformation. The group of vector gauge
transformations is Abelian, two such transformations with
parameters $\xi_1$ and $\xi_2$ combine to yield a transformation
with parameter $\xi_1 + \xi_2$. The two types of transformations
are independent of each other, and therefore combine as
\begin{eqnarray}
A' = UAU\dag + \phi\,,\qquad B' = UBU\dag + \D_{A'}\xi'\,,
\label{0701.combo1}
\end{eqnarray}
provided the one-form $\xi$ transforms homogeneously in the
adjoint as $\xi' = U\xi U\dag$.

The first question that needs to be asked concerns the uniqueness
of the transformation law in Eq.~(\ref{0701.ugauge}), in the light
of Eq.~(\ref{0701.vector}). How arbitrary is the one-form $\xi\,$?
It is clear that if the two types of transformations are to remain
independent, $\xi$ may not be a connection,  i.e., may not
transform as $\xi' = U\xi U\dag + \phi$. This is because
connections do not form a group under addition, for any two
one-forms $\xi_1$ and $\xi_2$ which transform like connections,
$\xi_1 + \xi_2$ does not transform like a connection. At least
$\xi$ cannot be an arbitrary connection one-form. But it is
possible to choose a connection, which depends on $U$, in place of
$\xi$ and construct novel symmetries by mixing the two types of
transformations.

There is a connection with some degree of arbitrariness which can
be used in place of $\xi$, or more precisely in place of $\xi'$ in
Eq.~(\ref{0701.combo1}). It is the flat connection $\phi$,
constructed from the gauge transformation $U$ as $\phi = - \D
UU\dag$. This choice modifies Eq.~(\ref{0701.combo1}) to
\begin{eqnarray}
A' &=& UAU\dag + \phi\,, \nonumber\\
B' &=& UBU\dag + \D_{A'}\phi\,\nonumber \\
&=& UBU\dag +  \Big(A'\W\phi + \phi\W A' + \D\phi\Big)
\,\nonumber \\ 
&=& UBU\dag +   \Big(UAU\dag\W\phi + \phi\W UAU\dag +
\phi\W\phi\,\Big)\,.  
\label{0701.gauge}
\end{eqnarray}
The only arbitrariness in these transformations is in $U$,
there is no arbitrary vector field, so this really has nothing to
do with the vector gauge transformations. In fact, it is easy to
see that this is nothing but a new gauge transformation law of
$B$. To see this, it is sufficient to show that two successive
gauge transformations of $B$ combine according to the group
multiplication law of SU(N).

Consider two gauge transformations $U_1$ and $U_2$, applied
successively to the fields. According to Eq.~(\ref{0701.gauge}),
the fields transform under $U_1$ as
\begin{eqnarray}
A_1 = U_1 A U_1\dag + \phi_1\,,\qquad
B_1 = U_1 B U_1\dag + \D_{A_1}\phi_1\,,
\label{0701.succ1}
\end{eqnarray}
with $\phi_1 = -\D U_1 U_1\dag\,,$ and then under $U_2$ as
\begin{eqnarray}
A' = U_2 A_1 U_2\dag + \phi_2\,,\qquad
B' = U_2 B_1 U_2\dag + \D_{A_2}\phi_2\,,
\label{0701.succ2}
\end{eqnarray}
with $\phi_2 = -\D U_2 U_2\dag\,.$ Substituting for $A_1$ and $B_1$
in Eq.~(\ref{0701.succ2}) their expressions from
Eq.~(\ref{0701.succ1}), I get back Eq.~(\ref{0701.gauge}), but with
$U = U_2 U_1$. Obviously, the transformation is invertible, since
the choice $U_2 = U_1\dag$ will give the identity transformation.
Taking $U\to 1$ continuously will also find the identity
transformation, which shows that the transformation for $B$ is
connected to the identity. So it is perfectly acceptable to treat
Eq.~(\ref{0701.gauge}) as the SU(N) gauge transformation law for
the fields.  However, the action is not exactly invariant under the
gauge transformation, but changes by a total divergence,
\begin{equation}
\int \Tr B'\W F'
= \int \Tr  B\W F +  \int  \D\,\Tr\,(\phi\W UFU\dag)\,,
\label{0701.deltaS}
\end{equation}
where I have used Bianchi identity, cyclicity of trace, and the
fact the $\phi$ is flat.

The gauge transformation law of $A$ is of course the standard one,
but that of $B$ is unusual in several respects. The fact that it is
inhomogeneous sets $B$ apart from all fields, other than the gauge
field $A$, which carry a representation of the gauge group. It
makes $B$ appear more like a connection than is usually
thought. For an Abelian gauge group, all commutators vanish, and
$B' = B$ is recovered, just as it would be for the homogeneous
transformation law of Eq.~(\ref{0701.ugauge}).  Note also that the
inhomogeneous transformation of $B$ in Eq.~(\ref{0701.gauge}) makes
sense only if $B$ is a two-form and $A$ is a one-form, and
therefore is a symmetry of the action only in four dimensions. It
is possible to construct similar transformations for higher
$p$-forms in $p+2$ dimensions. For example a three form $B_3$
coupled to the gauge field via a $B_3\W F$ interaction in five
dimensions can be taken to transform under the gauge group as
\begin{equation}
B'_3 = UB_3U\dag + UAU\dag\W\phi\W\phi - \phi\W\phi\W UAU\dag\,. 
\label{0701.B3gauge}
\end{equation}
The corresponding variation in the action is then
\begin{equation}
\delta\int\Tr B_3\W F = \int\D\Tr(\D\phi\W UFU\dag)\,.
\end{equation}
%
It is possible to consider other generalizations of the
transformation law. But three-forms or higher $p$-forms and
corresponding higher dimensional actions will not be explored in
this paper. Note that the specific transformation law of
Eq.~(\ref{0701.gauge}) holds only in four dimensions.

For results and theorems which use perturbation theory,
superficially this new transformation law does not pose a major
problem. This is because the inhomogeneous part is in some sense
strictly {\em finite}, becoming irrelevant for gauge
transformations infinitesimally close to the identity. But those
transformations are all that is needed for an analysis using the
Becchi-Rouet-Stora-Tyutin (BRST) differential. Let me write the
transformation law for $B$ with Lorentz indices and coupling
constants restored,
\begin{eqnarray}
B'_{\mu\nu} &=& U B_{\mu\nu} U\dag - \frac1m[UA_{[\mu} U\dag,
\partial_{\nu]} UU\dag] \,\nonumber \\
&& \qquad \qquad \quad + \frac{i}{gm}[\partial_\mu UU\dag,
\partial_\nu UU\dag]\,,  
\label{0701.gaugeindex}
\end{eqnarray}
where I have written $A_\mu = A^a_\mu t^a, B_{\mu\nu} =
B^a_{\mu\nu}t^a,$ etc. I have explicitly included the constant $m$,
but it seems to have any significance only if the two-form is
dynamical, with mass dimension one. The BRST transformations
corresponding to Eq.~(\ref{0701.gauge}) are
\begin{eqnarray}
s A^a_\mu &=&  \partial_\mu\omega^a +
gf^{abc}A^b_\mu\omega^c\,,\nonumber \\
s B^a_{\mu\nu} &=& gf^{abc} B^b_{\mu\nu}\omega^c +
\frac{g}{m}f^{abc}A^b_{[\mu}\partial_{\nu]}\omega^c\,,
\nonumber\\  
s \omega^a &=& -\half gf^{abc}\omega^b\omega^c\,.
\label{0701.BRST1}
\end{eqnarray}
This BRST operator is nilpotent, $s^2 = 0$, as a BRST operator
should be.  Comparison with the conventional BRST rules~\cite{%
Hwang:1997er,Lahiri:1997dm,Lahiri:2001uc,Cattaneo:1998eh,%
Thierry-Mieg:1983un} shows that the inhomogeneous part of the gauge
transformation law of $B$ is like the vector gauge transformation
with $\partial_\mu\omega^a$ playing the role of the vector
parameter. This is why known perturbative results, which includes
the vector gauge transformations in the analysis, may not need much
modification. But quite clearly it is not correct to think of the
new gauge transformation as a special case of vector gauge
transformation with parameter $\partial_\mu\omega^a$, since it is
not possible to start from the latter and exponentiate it to get
Eq.~(\ref{0701.gauge}).

The total divergence which appears in the variation of the action
will contribute to the conserved current of gauge symmetry.
Consider an SU(N) gauge transformation $U = 1 + ig\xi^a t^a$
infinitesimally close to the identity. The corresponding change in
the action is
\begin{equation}
\delta\int\frac {m}{4} \epsilon^{\mu\nu\rho\lambda}
B^a_{\mu\nu}F^a_{\rho\lambda} = -\frac{1}{2}\int
\epsilon^{\mu\nu\rho\lambda} \partial_\mu(\xi^a \partial_\nu
F^a_{\rho\lambda}). 
\label{0701.infdeltaS}
\end{equation}
It follows that the conserved current of gauge symmetry has a {\em
topologically conserved} component
\begin{eqnarray}
j^{a\mu}_{T} = -\frac{1}{2} \epsilon^{\mu\nu\rho\lambda}
\partial_\nu F^a_{\rho\lambda}\,.
\label{0701.jtopo}
\end{eqnarray}
This looks like a current of non-Abelian Dirac monopoles. This
current is not gauge-covariant, but in a configuration where $F$
vanishes on the boundary (e.g. Euclidean finite action) this makes
a vanishing contribution to the conserved charge.

\section{\label{actions}Symmetric actions}
In usual gauge theories, with usual objects, the gauge
transformation law of a field depends only on the transformation
$U$ and the field itself. The appearance of the gauge field $A$ in
the gauge transformation law for $B$ is thus quite unusual.
However, these are objects in a theory with local SU(N) symmetry.
If there is any dynamics of any field at all, not just dynamics of
$B$, the theory must contain gauge covariant derivatives, and
therefore a gauge field. Even if the gauge field to which $B$
couples is not the dynamical gauge field $A$, but a flat
connection, there is an inhomogeneous component of the gauge
transformation law of $B$. This situation arises for the naive
generalization of the duality relation between a two-form and a
scalar~\cite{Oda:1990tp,Smailagic:2000hr}, or outside the horizon
of black holes with a non-Abelian topological
charge~\cite{Lahiri:1992yz}. The inhomogeneous part of the gauge
transformation law for $B$ does not affect the results in these
cases.

If the gauge field $A$ is not constrained to be flat, it is
possible to construct interesting actions which are invariant under
the `new and improved' gauge transformation law.  Note that
\begin{equation}
\D A' = -\phi\W UAU\dag + U\D AU\dag - UAU\dag\W\phi + \D\phi\,,
\label{0701.gaugeAA}
\end{equation}
so that both the combinations $(B + \D A)$ and $(B -
 A\W A)$ transform covariantly under the gauge group,
\begin{eqnarray}
B' + \D A' &=& U(B + \D A)\,U\dag\,,\nonumber \\
B' -  A'\W A' &=& U(B -  A\W A)U\dag\,.
\label{0701.B-DA}
\end{eqnarray}
An interaction Lagrangian of the form $\Tr(B + \D A)\W F$ or $\Tr(B
- A\W A)\W F$ will be exactly invariant under gauge
transformations, and the conserved current for SU(N) gauge
transformations will be gauge covariant.
For either choice, the extra term is a total divergence,
\begin{eqnarray}
\Tr\,\D A\W F &=& \D\,\Tr (A\W\D A + \frac13 A\W A\W A)\,,
\nonumber \\ 
\Tr\,A\W A\W F &=& \D\,\Tr\, \frac13 A\W A\W A\,.
\label{0701.AAA}
\end{eqnarray}

I can now construct several novel actions involving only $A$ and
$B$, with interesting physical implications, by demanding
invariance only under the gauge transformation of
Eq.~(\ref{0701.gauge}). I shall ignore vector gauge transformations
in this section, coming back to them in the next section. One
reason for this exercise is to show that an inhomogeneous
transformation law for $B$ does not automatically rule out
construction of invariant actions. Another reason will become
apparent in the next section where I will argue that some of the
actions constructed below are gauge fixed versions of actions with
vector gauge symmetry. The starting point is either $\Tr B\W F$,
which is invariant up to a total divergence, or $\Tr(B + \D A)\W
F$, which is exactly invariant, but CP-violating.  To either of
this a quadratic term, constructed out of $(B + \D A)$, or the
covariant field strength $(\D_A B - \D F)$, can be added for an
invariant action.

The original topological action itself is of course the simplest
example of an action symmetric under the new SU(N) gauge
transformation rules. This action can be obviously modified without
changing its classical content, by replacing $B$ by $(B + \D A)$
or $(B - A\W A)$, to give a triplet of actions differing only by a
total divergence,
\begin{eqnarray}
S_0 &=& \int \Tr B\W F\,\nonumber \\
    &=& \int \Tr(B + \D A)\W F\,\nonumber \\
    &=& \int \Tr(B - A\W A) \W F \,,
\label{0701.topo1}
\end{eqnarray}
where the equalities hold up to total divergences. Although the
equations of motion derived from these actions are the same,
quantum theories built from them will behave differently in the
non-perturbative regime. They also have different properties under
$CP$. To see this, note that $B$ can be chosen to transform under
$CP$ like $*F$ so that the first action can be made
$CP$-conserving.  But the other two equivalent actions will
necessarily violate $CP$.

Another class of $CP$-violating actions come from using quadratic
terms of the type $B\W B$. An example is the action
\begin{eqnarray}
S_1 &=&   \int \Tr \Big( B\W F - \half (B + \D A)\W(B + \D A)
\Big) \nonumber \\
    &=&   \int  \Tr \Big(B\W A\W A - \half B\W B - \half \D A\W \D A
\Big).\;\;
\label{0701.BWB}
\end{eqnarray}
The equation of motion following from this action by varying $B$ is
$B = A\W A\,.$ This may be substituted in the
action~\cite{Henneaux:1991my}, giving
\begin{eqnarray}
S_1 &=& \int \Tr \Big( A\W A\W F - \half F\W F \Big)\,\nonumber \\
    &=& \int -\half\,\Tr\,\D A\W \D A\,,
\label{0701.FWF}
\end{eqnarray}
which is a total divergence. Obviously there are many variations on
this theme which arise from replacing $B$ in $B\W F$ by either of
the two gauge covariant combinations and from using $(B - A\W A)$
in one or both factors of the second term.  All these actions are
classically equivalent, i.e., they differ by total divergences.
They are also equivalent in the path integral in the same sense,
after Gaussian integration over $B$. So all the information about
the corresponding quantum theories reside on the boundary of
space-time. An interesting point is that some actions of this
$CP$-violating class vanish altogether upon using the equations
of motion for $B$. An example is
\begin{equation}
S'_1 = \int\Tr \left( (B - A\W A)\W F - (B + \D A)\W (B - A\W
A)\right)\,,
\label{0701.nullaction}
\end{equation}
for which the equation of motion of $B$ is $B = A\W A\,,$ so that
$S'_1 = 0\,.$

An example of an action which is not a total divergence is a first
order formulation of Yang-Mills theory, along the lines
of~\cite{Cattaneo:1998eh,Fucito:1997sq,Accardi:1997bf}, but without
vector gauge symmetry,
\begin{eqnarray}
S_2 = \int\Tr \Big(B\W F + \half (B + \D A)\W*(B + \D A)  \Big)\,.
\label{0701.BFYM1}
\end{eqnarray}
The equation of motion for $B$ is $B + \D A = *F$, which can be put
back into the action. This action is then classically equivalent,
i.e. equal up to a total divergence, to Yang-Mills theory. Again,
there are several variations on this theme, not all of which have
local dynamics. One example of this type is
\begin{equation}
S'_2 = \int\Tr \Big(B\W F + (B + \D A)\W*(B - A\W A)  \Big)\,.
\label{0701.BFYM2}
\end{equation}
The equation of motion for $B$ derived from this action is $B =
\half(-\D A + A\W A + *F)\,.$ Substituting this expression for $B$
into the action produces
\begin{eqnarray}
S'_2 = \int-\half\Tr\,  \D A\W \D A \,,
\label{0701.BFYMtd}
\end{eqnarray}
which is a total divergence. So even though the action written in
terms of $B$ seems to require a metric, actually the dynamics it
describes is independent of the metric and lives fully on the
boundary of the space-time.

I should mention actions of another type before ending this
section. Since the field strength $F$ is a gauge covariant object,
any constant multiple of of $F$ can be added to $(B - A\W A)$ to
produce a gauge covariant combination. Among all such combinations,
$(B - A\W A - F)$ is somewhat special. It will be argued in the
next section that actions constructed out of this combination of
fields are gauge equivalent to actions with explicit vector gauge
symmetry.  Two examples are the first order formulations of the
Yang-Mills action and the topological Yang-Mills action,
\begin{eqnarray}
S_{YM}  &=& \int\Tr\Big[(B - A\W A -F)\W F \,\nonumber \\
&&\quad + \half (B - A\W A -F)\W *(B - A\W A -F)
\Big]\,\nonumber \\  
        &\equiv& \int \half\Tr\,F\W *F\,, \\
S_{TYM} &=& \int\Tr\Big[(B - A\W A -F)\W F \,\nonumber \\
&&\quad - \half (B - A\W A -F)\W (B - A\W A -F)\Big] \,\nonumber \\
        &\equiv& \int \half\Tr\,F\W F\,,
\label{0701.YM-TYM}
\end{eqnarray}
where the equivalences are obtained by substituting the equations
of motion for $B$ into the respective actions. 

Another action which should be mentioned is that of a dynamical
non-Abelian two-form. Dynamics requires a gauge covariant field
strength to be defined for $B$, and the purpose is served by
$\tilde H = \D_A(B + \D A) \equiv \D_A(B - A\W A) = \D_A B - \D F.$
Like other gauge covariant combinations mentioned in this section,
this field strength is not invariant under vector gauge
transformations. With the help of this field strength, I can now
write down an action in which the two-form $B$ is a dynamical
field,
\begin{eqnarray}
S_3 = \int\Tr\Big(\half \tilde H\W*\tilde H + \half F\W*F + B\W F
\Big)\,.
\label{0701.dynB1}
\end{eqnarray}
The $B$-independent part of this action is a nonlinear
$\sigma$-model for the gauge field $A$. The term quadratic in
derivatives of $A$ reads, with gauge and Lorentz indices restored, 
\begin{eqnarray}
-\frac14(\delta^{bd} + g^2 f^{abc}f^{ade}A^c_\lambda A^{\lambda e})\, 
\partial_{[\mu}A^b_{\nu]}\partial^{[\mu}A^{\nu]d}\,\nonumber \\
 - \frac12 g^2 f^{abc}f^{ade} \partial_{[\mu}A^b_{\nu]}A^c_\lambda
\partial^{[\nu}A^{\lambda]d}A^{\mu e}\,. 
\label{0701.nlsA}
\end{eqnarray}
This action is not power counting renormalizable, nor should it
lead to a consistent quantum theory, like any nonlinear
$\sigma$-model in four dimensions. But I will argue in the next
section that this action is gauge equivalent to the theory with an
auxiliary field and explicit vector gauge symmetry, which is known
to be renormalizable~\cite{Lahiri:2001uc}. In the Abelian limit
where the structure constants vanish, $\tilde H$ becomes the usual
field strength for a set of Abelian two-forms, $\tilde H \to \D
B$. So this action then becomes the same as that for a set of
topologically massive Abelian gauge fields.

Obviously, many other actions can be constructed using
Eq.~(\ref{0701.gauge}) as the gauge transformation laws and they
will correspond to different physical systems. One can even
construct actions which do not have the $B\W F$ term as the
cornerstone. I will not explore such actions here. Finally, for
$p$-forms in $p+2$ dimensions, a generalized SU(N) gauge
transformation law can be trivially constructed by demanding
covariance of 
\begin{equation}
B_p - A\W A\W\cdots\W A\,\, (p\; factors)\,.
\end{equation}
Just as the result for $p=3$ differs from the one given earlier in
Eq.~(\ref{0701.B3gauge}), clearly there are several different
generalizations for higher $p$-forms.

\section{\label{aux}Auxiliary one-form}
The actions mentioned in the previous section, except for the pure
$B\W F$ action and its modifications as shown in
Eq.~(\ref{0701.topo1}), were not symmetric under vector gauge
transformations. These need to be reintroduced into the
discussion. The interaction Lagrangian $\Tr B\W F$ is symmetric up
to a total divergence under the vector gauge transformations of
Eq.~(\ref{0701.vector}). Any term quadratic in $B$, including a
possible kinetic term, is not invariant. This is obvious for terms
like $B\W B$ or $B\W *B$, for kinetic terms this is because the
`field strength' $\D_A B$ changes under these transformations,
\begin{eqnarray}
\D_A B \to \D_A B + F\W\xi - \xi\W F\,.
\label{0701.deltaH1}
\end{eqnarray}
Indeed a theorem~\cite{Henneaux:1997mf} asserts that a kinetic term
for the two-form, invariant under both types of gauge
transformations, cannot be constructed unless additional fields are
introduced to compensate for the vector gauge transformations. Note
that since this theorem was proven by using the BRST structure of
these theories, the modification of the gauge symmetry displayed in
the previous section should not change its proof.

Some actions which use auxiliary one-form fields to compensate for
the vector gauge transformations have been known for some
time~\cite{Lahiri:1992hz,Cattaneo:1998eh}. In these
actions, a one-form field $C$ is introduced, and is assumed to
shift under these transformations,
\begin{eqnarray}
A' = A\,,\qquad B' = B + \D_A\xi\,,\qquad C' = C + \xi\,.
\label{0701.vector2}
\end{eqnarray}
Obviously the combination $(B - \D_A C)$ remains invariant under
these transformations, as does the compensated field strength 
\begin{equation}
H = \D_A B - F\W C + C\W F\,. 
\label{0701.Hdef}
\end{equation}
How does $C$ behave under usual gauge transformations? Clearly it
has to transform in the adjoint representation. There is now no
need for $\D A$ to cancel the inhomogeneous part of gauge
transformations of $B$. Instead, the auxiliary field $C$ can be
taken to transform inhomogeneously, like a connection, under the
gauge group,
\begin{eqnarray}
A' &=& UAU\dag + \phi\,,\nonumber \\
B' &=& UBU\dag + \D_{A'}\phi\,,\nonumber \\
C' &=& UCU\dag + \phi\,.
\label{0701.ctrans}
\end{eqnarray}

With this choice, the combination $(B - \D_A C)$ transforms
covariantly under the gauge group, $(B - \D_A C) \to U(B - \D_A
C)U\dag\,.$ The field strength $H$ also transforms covariantly, $H
\to UHU\dag\,.$ Both these combinations, $(B - \D_A C)$ and $H\,,$
are also invariant under the vector gauge transformations in
Eq.~(\ref{0701.vector2}), with $\xi$ transforming homogeneously in
the adjoint, $\xi' = U\xi U\dag\,.$ Therefore, terms quadratic in
$(B - \D_A C)$ or $H$ can be used for construction of symmetric
actions, invariant under both usual and vector gauge
transformations. I will not mention such actions separately here,
they are discussed in~\cite{Lahiri:1992hz,Hwang:1997er,
Lahiri:1997dm,Lahiri:2001uc,Cattaneo:1998eh,Fucito:1997sq,
Accardi:1997bf}. Note that if $C$ is taken to transform
homogeneously under the gauge group, invariant actions have to be
constructed with $(B + \D A - \D_A C)$ or with $(B - A\W A - F -
\D_A C) \equiv B - \D_A(A + C)$. These possibilities will be
ignored, since they correspond to redefinitions of $C$.
Henceforth $C$ will be taken to transform like a gauge field, i.e.,
according to equation Eq.~(\ref{0701.ctrans}).

Since the flat connection $\phi$ does not transform homogeneously
under the gauge group, but the vector parameter $\xi$ has to do so,
it is clear that Eq.~(\ref{0701.ctrans}) cannot be a special case
of vector gauge transformations, despite the formal similarity. Nor
is it possible to take the vector parameter $\xi$ to transform like
a connection, because connections do not add, so vector gauge
transformations will not form a group. Alternatively, if the vector
parameter is an arbitrary connection, the properties of $C$ under
usual gauge transformations will be ill-defined.  This also provides
an additional reason why it is not possible to set $C=0$ by a gauge
choice. Since $C$ transforms like a connection, even if it is made
to vanish in one gauge, it will be non-zero upon an SU(N) gauge
transformation.

The BRST transformations for $C$ are similar to vector gauge
transformations as was the case for the two-form,
\begin{eqnarray}
s C^a_\mu = \frac1m\partial_\mu\omega^a + gf^{abc}C^b_\mu\omega^c\,.
\end{eqnarray}
Taken together with Eq.~(\ref{0701.BRST1}), the BRST operator is
nilpotent, $s^2 = 0\,,$ and all calculations which use BRST
analysis will go through.  In other words, the algebraic proofs of
renormalizability, for the massive vector
theory~\cite{Lahiri:2001uc}, or for the first-order Yang-Mills
theory~\cite{Fucito:1997sq, Accardi:1997bf}, should hold with minor
modifications.  For finite transformations, the picture differs
radically from that in those articles, $C$ is now a connection
under the gauge group, and a vector gauge transformation shifts it
by the difference of any two connections.

This picture leads to another interesting symmetry transformation.
Let $\tilde U(x)$ be an arbitrary SU(N) matrix valued field,
unrelated to the SU(N) gauge transformation, and let the flat
connection $\tilde\phi$ be defined as $\tilde\phi = -\D\tilde
U\tilde U\dag$.  A vector gauge transformation can be made upon the
fields $B$ and $C$ using a constant multiple of the difference of
$A$ and $\tilde\phi$. That may not seem particularly significant,
since it is just a special choice of the vector parameter. However,
if the theory contains only the field strength $H$ and not the
combination $(B - \D_A C\,)$, I can freely shift $B$ by a constant
multiple of $F$, the field strength of the gauge field, since under
this shift
\begin{equation}
B\to B + \alpha F\,,\qquad H \to H + \alpha\D_A F = H\,,
\label{0701.shiftH}
\end{equation}
by Bianchi identity. The interaction term $B\W F$ changes by a
total divergence under this shift, so this is a symmetry of the
action. For an Abelian two-form this shift $B \to B + \alpha F$ is
just a special case of the vector gauge transformations with the
special choice $\xi = A$. On the other hand, in the case of the
non-Abelian two-form, there is no choice of $\xi$ for which $F =
\D_A\xi\,.$ Indeed, such a shift for the non-Abelian two-form is
not even a local symmetry transformation, cannot be implemented by
a local constraint, and can be included in the BRST charge only by
using a constant ghost field~\cite{Lahiri:2001uc}.

Because of this, if the local transformation is combined with the
global shift, it produces a completely new type of symmetry
transformation,
\begin{eqnarray}
C' &=& C +\alpha (A - \tilde\phi)\,, \quad A' = A \,,\nonumber \\
B' &=& B +\alpha (A - \tilde\phi)\W(A - \tilde\phi) \,,
\label{0701.special}
\end{eqnarray}
with $\alpha$ an arbitrary constant. It is easy to see that the
SU(N) transformation properties of these fields are not affected
provided $\tilde\phi$ is taken to transform as a connection as
well. This changes the $B\W F$ interaction term by a total
divergence, and leaves the compensated field strength $H$ of
Eq.~(\ref{0701.Hdef}) invariant,
\begin{eqnarray}
\delta \int\Tr B\W F &=& \alpha \int\D\,\Tr (\frac13 A\W A\W A -
\tilde\phi\W F)\,,\nonumber \\ 
\delta(B - \D_A C) &=& - \alpha F\,,\nonumber \\
\delta H &=& 0\,.
\label{0701.nonpert}
\end{eqnarray}
It follows that an action containing the field strength $H$ and the
interaction $B\W F$ will be invariant (up to a total divergence)
under this set of transformations. Even the pure $B\W F$ action is
invariant under it. The corresponding conserved current, from the
variation of the $B\W F$ term, depends on the arbitrary flat
connection $\tilde\phi\,.$

Why is this a new type of symmetry? Since $\tilde U$ is an
arbitrary local SU(N) matrix, the flat connection $\tilde\phi$ is
constructed out of arbitrary space-time dependent parameters and
their derivatives. So these would seem to be local
transformations. However, in general local transformations are
generated by local first-class constraints. The vector gauge
transformations of Eq.~(\ref{0701.vector2}) are local, and they are
generated by local constraints~\cite{Lee:1998qu,Harikumar:2001eb}.
On the other hand, the shift $B \to B + \alpha F$ is a {\em global}
transformation for a non-Abelian $B$. But it is not possible to
write local constraints which can generate the transformations of
Eq.~(\ref{0701.special}). The simplest way to show this is by
noting that the local part of these transformations is not
connected to the identity, i.e., $\tilde\phi \to 0$ or $\tilde U
\to 1$ in Eq.~(\ref{0701.special}) does not lead to the identity
transformation. So in particular transformations for infinitesimal
$\tilde\phi$, i.e. for $\tilde U$ infinitesimally close to the
identity, cannot be written as transformations which are themselves
infinitesimally close to the identity transformation. Since the
Poisson brackets of a constraint with the variables of the theory
produce infinitesimal transformations, these transformations cannot
be produced by constraints. It also means that this symmetry will
not appear in the BRST charge.

This can be seen directly in the BRST approach, starting with the
transformation for infinitesimal $\tilde\phi$. For $\tilde U = 1 +
ig\dl\theta^a t^a\,,$ with $\dl$ an anticommuting constant and
$\theta^a$ an anticommuting field (the would be ghost), I can write
the infinitesimal changes in the fields as following from
Eq.~(\ref{0701.special}) as
\begin{eqnarray}
\delta B &=& \alpha A\W A + \alpha\dl (\D\theta\W A + A\W\D\theta)\,,
\nonumber \\ 
\delta C &=& \alpha (A + \dl\D\theta)\,,
\label{0701.spsmall}
\end{eqnarray}
where $\D\theta = igt^a\partial_\mu\theta^a\D x^\mu\,.$ Obviously
the derivative $\delta/\dl$, which would be part of the BRST
operator, does not have any meaning. This is because $\alpha$ has
been treated as a finite constant, so perhaps it should be replaced
by $\dl\,\alpha$, where $\alpha$ is now a constant ghost, as
in~\cite{Lahiri:2001uc}? This will clearly not produce the correct
BRST operator in this case either, because then the
$\theta$-dependent terms will disappear from
Eq.~(\ref{0701.spsmall}). So there is no BRST construction which
includes this symmetry, which is another way of showing that it
cannot be generated by local constraints, since the BRST charge has
to include all local constraints. Since the BRST charge is
fundamental to the local Hamiltonian quantization of gauge theories
(see for example~\cite{Henneaux:1992ig, Barnich:2000zw}), another
possible interpretation of this is that a full quantum theory of
this system must necessarily include non-local objects
(i.e. strings) and operators, with appropriate induced actions of
the symmetry group, in order to maintain the classical
symmetries. This is not very surprising since a two-form naturally
couples to world sheets.

This conclusion is based on the failure to assign a non-zero ghost
number to an infinitesimal $\tilde\phi\,.$ On the other hand it is
possible to construct a BRST operator which includes an arbitrary
flat SU(N) connection $\tilde\phi\,$ of zero ghost number and a
constant field $\alpha$ of ghost number one. Then there is a family
of BRST operators parametrized by $\tilde\phi\,,$ for a given set
of fields and ghosts, including $\alpha\,.$ Any two such BRST
operators anticommute, suggesting a degeneracy of quantum states,
labeled by flat connections. Lifting this degeneracy will require
choosing a specific flat connection. This is analogous, but not
identical, to choosing a vacuum in spontaneously broken gauge
theories.  The transformations of Eq.~(\ref{0701.special}) should
not therefore be confused with usual local symmetry
transformations. These should be called {\em semiglobal}
transformations, elements of a class of global transformations
parametrized by local SU(N) matrices $\tilde U$.

The fact that the auxiliary connection $C$ can be shifted by the
difference of two connections also provides a way of relating
actions constructed in this section with the help of $C$, and those
of the previous section, constructed without $C$ and without vector
gauge symmetry. Consider the case where these two connections are
$C$ and $A$, i.e., consider a vector gauge transformation with $\xi
= \alpha(A - C)\,$ where $\alpha$ is some constant. The transformed
fields are
\begin{eqnarray}
B' &=& B + \alpha F + \alpha A\W A - \alpha\, \D_A C\,\nonumber \\
C' &=& (1 - \alpha)C + \alpha A\,.
\label{0701.CAtrans}
\end{eqnarray}
The compensated field strength $H$ remains invariant, as it should
under a vector gauge transformation, but in addition it has a
familiar form for $\alpha = 1\,$ in terms of the transformed
fields. If I choose $\alpha = 1\,,$ in Eq.~(\ref{0701.CAtrans}),
the two connections become related to each other by $C' = A\,,$ and
then the field strength is $H = \D_A B' - F\W A + A\W F$, which
because of Bianchi identity can be written as $H = \D_A B' - \D
F$. This has the same form as the field strength defined in the
previous section, before the introduction of the auxiliary
one-form.  So even though $C$ cannot be set to vanish by a gauge
choice in Eq.~(\ref{0701.Hdef}), it can be absorbed into $B$ in the
aforementioned sense, whereby it is `replaced' by $A\,.$ Note also
that for $\alpha = 1\,,$ the invariant combination $(B - \D_A C)$
can be written as 
\begin{equation}
B' - \D_{A'}C' = B' - \D_A A = B' - F - A\W A\,.
\end{equation}
This links the actions mentioned in the previous section with those
mentioned here.  Finally, note that by a similar argument, $C$ can
also be `replaced' by an arbitrary flat connection $\phi$ via a
vector gauge transformation.

\section{\label{twocon}Two connections for two-form}
In the previous section it was shown that the auxiliary vector
field $C$ transforms like a gauge field under usual gauge
transformations, and shifts by the `difference of two connections'
under a vector gauge transformation. It was also shown that by an
appropriate vector gauge transformation, the auxiliary field could
be `shifted away' to be replaced, in a manner, by the gauge field
$A$. In this section the picture of $C$ as a second connection in
the theory will be made more explicit, and more actions will be
shown to be related by gauge symmetry to the ones already
mentioned.

The starting point is the observation that the auxiliary connection
$C$ always appears in conjunction with the two-form $B$. In fact,
it is needed for all actions with vector gauge symmetry, except for
the pure $B\W F$ action. But even this form of the action need not
be treated as sacrosanct, since $C$ can be absorbed in $B$ in a
specific manner using vector gauge transformations. Then I can
conclude that the non-Abelian two-form $B$ and the auxiliary
connection $C$ cannot be separately included in any theory, but has
to be considered as a pair. 

Is it then possible to formulate the gauge transformation laws of
the pair $(B, C)$ purely in terms of themselves without referring
to the usual gauge field $A$ as was done earlier? This would remove
the dependence of $B$ on the gauge field $A$, which is somewhat
artificial, since there is no converse dependence --- a theory of
the gauge field can be defined without invoking $B$. This requires
that the gauge transformation laws described in
Eq.~(\ref{0701.ctrans}) be thought of as being in a special choice
of gauge for the vector gauge transformations.

Following this argument, let me first rewrite the SU(N) gauge
transformations for $A, B\,,$ and $C$ as
\begin{eqnarray}
A' &=& UAU\dag + \phi\,\nonumber \\
B' &=& UBU\dag + UCU\dag\W\phi + \phi\W UCU\dag  + \phi\W\phi\,
\nonumber \\
C' &=& UCU\dag + \phi\,.
\label{0701.ctrans2}
\end{eqnarray}
These are the same as the rules of Eq.~(\ref{0701.ctrans}), but
with the gauge field $C$ taking the place of the gauge field $A$ in
the transformation rule for $B$. It will be shown later that these
are in fact equivalent to the earlier rules.

Just as in the earlier incarnation, these transformations combine
according to the group multiplication law of SU(N). This can be
seen by applying two successive SU(N) gauge transformations $U_1$
and $U_2$ and using Eq.s~(\ref{0701.succ1}) and (\ref{0701.succ2}),
but substituting $C$ for $A$ in those equations. One immediate
consequence of choosing Eq.~(\ref{0701.ctrans2}) as the gauge
transformation rules is that the $B\W F$ term is no longer gauge
invariant, as expected from the discussion above. The action that
should be used in its place is 
\begin{equation}
S_0 = \int\Tr\,(B + \D C)\W F = \int\Tr\,(B\W F - C\W\D
F)\,, 
\label{0701.topo2}
\end{equation}
where the second equality holds up to a total divergence.  The
second term of this action looks like a magnetic monopole term in
the Abelian limit $g\to 0\,.$ It does not have a clear
interpretation for non-Abelian theories, since $\D F$ is not a
gauge-covariant object.

In any case, this choice of gauge transformation rules simplifies
the vector gauge transformations quite remarkably. Now it is the
combination $B + \D C$ which transforms homogeneously, $B' + \D C'
= U(B + \D C)U\dag$. I can now define vector gauge transformations
for the non-Abelian two-form without reference to any connection
and in fact these are exactly the same as the familiar Kalb-Ramond
symmetry~\cite{Kalb:1974yc}
\begin{eqnarray}
B \to B + \D\xi\,,\qquad C \to C - \xi\,.
\label{0701.KR}
\end{eqnarray}
Here $\xi$ is a one-form which transforms homogeneously under the
gauge group, i.e., the difference of two connections, $\xi \to U\xi
U\dag\,.$ This choice ensures that $C'$ also transforms like a
connection, and the behavior of $B$ under SU(N) gauge
transformations in maintained. Indeed, if an SU(N) gauge
transformation $U$ is applied subsequent to the Kalb-Ramond
transformation, and $\xi$ is taken to transform homogeneously, the
fields transform as
\begin{eqnarray}
C' &=& U(C - \xi)U\dag + \phi\,\nonumber \\
B' &=& U(B + \D\xi)U\dag + UC'U\dag\W\phi + \phi\W UC'U\dag +
\phi\W\phi\,. \nonumber \\
\label{0701.KRgauge}
\end{eqnarray}
Therefore, just as in the previous section, the auxiliary one-form
$C$ transforms like a gauge field under ordinary SU(N) gauge
transformations, and shifts by the difference of two connections
under Kalb-Ramond symmetry. Note that I am making a distinction
between the vector gauge symmetry of Eq.~(\ref{0701.vector2}),
which involves the gauge covariant exterior derivative $\D_A\,,$
and the Kalb-Ramond symmetry, which requires only the ordinary
exterior derivative $\D\,.$ I will show that actions invariant
under the gauge transformations of Eq.~(\ref{0701.ctrans2}) and the
Kalb-Ramond transformations are also equivalent to the actions of
section~\ref{actions} by a symmetry transformation.

Before going on to discuss invariant actions, let me briefly
mention one peculiarity of the $(B, C)$ system. The combination $(B
+ \D C)$ appears to have another obvious symmetry, under $C \to C +
\D\chi$ where $\chi$ is a scalar. This symmetry is not compatible
with gauge symmetry, since $C + \D\chi$ cannot transform as in
Eq.~(\ref{0701.ctrans2}) for any choice of $\chi$. So this symmetry
is not likely to have any physical significance. However, the field
$B$ and $C$ must always appear in the combination $(B + \D C)$ and
its derivatives, so this symmetry will always be present in the
action.

Invariant actions involving the $(B, C)$ pair are now easy to
construct. Note that a `field strength' for $C$, 
\begin{equation}
F_C = \D C + C\W C\,,
\label{0701.FC}
\end{equation}
is covariant under SU(N) gauge transformations, but is not
invariant under Kalb-Ramond transformations. So it will not appear
in an invariant action. Note also that even though $C$ transforms
like a gauge field, the gauge covariant derivative is still defined
as $\D_A\,,$ since it is a good idea to leave the covariant
derivative unmolested after a Kalb-Ramond transformation.

Then the actions invariant under both the gauge group and the
Kalb-Ramond symmetry are already known. These are simply the
actions mentioned in section~\ref{actions}, but with $(B +\D C)$
replacing $(B + \D A)$. The $B\W F$ term, or its equivalents as in
Eq.~(\ref{0701.topo1}), has to be replaced by the term $S_0$ as in
Eq.~(\ref{0701.topo2}) as well. For example, the action
corresponding to the parity violating action $S_1$ of
Eq.~(\ref{0701.BWB}) is
\begin{eqnarray}
S_1 &=& \int \Tr \Big( (B + \D C)\W F - \half (B + \D C)\W(B +
\D C) \Big) \nonumber \\
    &=&   \int \half\, \Tr \, F\W F\,,
\label{0701.BWB2}
\end{eqnarray}
where the second equality comes from substituting the equation of
motion $B = F - \D C$ into the action.  Another example is the
action for the first order formulation of Yang-Mills theory, which
is now
\begin{eqnarray}
S_2 &=& \int\Tr \Big((B +\D C)\W F + \half (B + \D C)\W*(B + \D
C)  \Big)\,\nonumber \\
    &=& \int \half\,\Tr\,F\W *F\,. 
\label{0701.BFYM3}
\end{eqnarray}
Again I have substituted the equation of motion for $B$, which is
$B = *F - \D C\,,$ into the action to produce the second equality.
Unlike in section~\ref{actions} where actions were constructed from
only the gauge field $A$ and the two-form $B$, and only invariance
under SU(N) gauge transformations was imposed, the requirement of
Kalb-Ramond symmetry rules out various combinations. For example,
the combination $(B - C\W C)\,,$ while covariant under SU(N) gauge
transformations, is not invariant under the Kalb-Ramond
transformation. On the other hand, just as in
section~\ref{actions}, any constant multiple of the gauge field
strength $F$ can be added to $(B +\D C)$ for a gauge covariant,
Kalb-Ramond invariant combination. It is easy to see that actions
built with those combinations are equivalent to the ones already
mentioned.

As mentioned earlier, the gauge covariant derivative, which is
needed to construct dynamical actions, should be taken to be
$\D_A\,,$ even though $C$ is also a gauge field, because $C$ is not
invariant under Kalb-Ramond transformations. The field strength for
the two-form is constructed with the gauge covariant derivative
$\D_A$,
\begin{equation}
H = \D_A(B + \D C) = \D_A B + A\W\D C - \D C\W A\,,
\label{0701.Hfinal}
\end{equation}
and the action for the dynamical two-form is then
\begin{equation}
S_3 = \int\Tr\Big(\half H\W* H + \half F\W*F + B\W F
- C\W\D F\Big)\,.
\label{0701.dynBC}
\end{equation}
Needless to say, this action reduces to the usual Abelian action of
topologically massive fields (either $A$ or $B$) in the Abelian
limit. Another interesting point is that in that limit this action
can be thought of as a (partial) first order formulation, with $C$
being a dual gauge field.

In the non-Abelian theory, $C$ is shifted by the difference of two
connections under a Kalb-Ramond transformation. I can choose the
two connections to be $C$ and $A$ as in the section~\ref{aux}, $\xi
= \alpha(C - A)\,.$ Then
\begin{eqnarray}
B' &=& B + \alpha\D(C - A)\,,\nonumber \\
C' &=& C - \alpha (C - A)\,.
\label{0701.CAtrans2}
\end{eqnarray}
Since this is a special case of the Kalb-Ramond transformations,
the combination $(B +\D C)$ remains invariant, as does the field
strength $H$. For $\alpha = 1\,,$ this has the effect of replacing
$C$ by $A$. Thus the actions mentioned earlier in
section~\ref{actions} are equivalent to the actions mentioned in
this section as well.

The field strength $H$ of Eq.~(\ref{0701.Hfinal}) is invariant, as
before, under a semiglobal transformation, constructed with the
help of an arbitrary flat connection $\tilde\phi$ transforming in
the adjoint representation of the gauge group. But these now take a
slightly different form,
\begin{eqnarray}
B' &=& B + \alpha\,(A\W A - \tilde\phi\W\tilde\phi)\,,\nonumber \\
C' &=& C - \alpha\,(A - \tilde\phi)\,.
\label{0701.semiglob2}
\end{eqnarray}
The combination $(B +\D C)$ is not invariant under these
transformations, while the field strength $H$ is. As before, these
are not connected to the identity transformation for $\tilde\phi\to
0\,.$ The conserved current for this transformation depends on the
form of the interaction term. For the action of
Eq.~(\ref{0701.dynBC}) the current depends on the flat connection
$\tilde\phi$ and is proportional to 
\begin{equation}
\Tr\,(A\W A\W A + \tilde\phi\W F)
\end{equation}
As in section~\ref{aux}, the BRST operators which produce the
transformations of Eq.~(\ref{0701.semiglob2}) must be constructed
with a constant ghost field $\alpha$ and an arbitrary (commuting)
flat connection $\tilde\phi$. Consequently they are again
parametrized by the space of flat connections $\tilde\phi\,.$

\section{\label{last}Summary of results}
In this paper I have argued that an antisymmetric tensor potential
valued in the adjoint representation should have an inhomogeneous
component in its gauge transformation rule, as shown in
Eq.~(\ref{0701.gauge}). It is not clear if there is a geometrical
interpretation of this rule, i.e. a geometrical object which
corresponds to this. But at any rate it opens up new avenues of
investigation. These new gauge transformation rules are not in
contradiction with theorems based on BRST analysis. In particular
the new rules do not obviate the need for an auxiliary one-form in
actions containing terms quadratic in $B$.

I have also shown that the auxiliary one-form $C$ transforms as a
gauge field. Therefore it cannot be shifted away to zero by a
vector gauge transformation, because an ordinary gauge
transformation changes a vanishing gauge field to a non-vanishing
flat connection. But it can be `replaced' by the usual gauge field
$A$ via a vector gauge transformation. The actions in
section~\ref{aux} can be related in this way to the actions in
section~\ref{actions}, which may be thought of as being somewhat
analogous to a {\em unitarity gauge} choice for these theories. An
outcome of this is that the action for the topological mass
generation mechanism becomes gauge equivalent to a nonlinear
$\sigma$-model for the gauge field, as in Eq.~(\ref{0701.nlsA}).
This action also has a new kind of {\em semiglobal} symmetry which
depends on arbitrary flat connections, but is connected to the
identity only in the limit of a vanishing global parameter. This
symmetry is not generated by local constraints, unlike all other
known local symmetries in classical field theory. The conserved
current or this symmetry is parametrized by the space of flat
connections. It is in fact possible to construct a family of BRST
operators, parametrized by flat connections, which anticommute with
one another.

I have also shown that it is possible to define inhomogeneous gauge
transformation rules for the pair of fields $(B, C)$ without
referring to the gauge field $A$. The actions invariant under these
rules are symmetric under the usual (Abelian) Kalb-Ramond symmetry.
Unlike for the vector gauge symmetry, the integral of $B$ on a
closed 2-surface is invariant under Kalb-Ramond symmetry. The
actions of section~\ref{twocon} are also equivalent, by gauge
transformations and field redefinitions, to the actions discussed
in the earlier sections. The symmetries and actions mentioned in
this paper should be useful for all $B\W F$ type theories,
including gravity.








\begin{thebibliography}{99}

\bibitem{Freedman:1977pa}
D.~Z.~Freedman,
{\sl ``Gauge Theories Of Antisymmetric Tensor Fields,''}
Cal Tech preprint CALT-68-624.
\bibitem{Seo:1979id}
K.~Seo, M.~Okawa and A.~Sugamoto,
Phys.\ Rev.\ D {\bf 19}, 3744 (1979).
\bibitem{Freedman:1981us}
D.~Z.~Freedman and P.~K.~Townsend,
Nucl.\ Phys.\ B {\bf 177}, 282 (1981).

\bibitem{Schwarz:1978cn}
A.~S.~Schwarz,
Lett.\ Math.\ Phys.\  {\bf 2}, 247 (1978);

\bibitem{Schwarz:1979ae}
A.~S.~Schwarz,
Commun.\ Math.\ Phys.\  {\bf 67}, 1 (1979).

\bibitem{Witten:1988ze}
E.~Witten,
Commun.\ Math.\ Phys.\  {\bf 117}, 353 (1988).

\bibitem{Birmingham:1991ty} 
D.~Birmingham et al,
Phys.\ Rept.\  {\bf 209}, 129 (1991).

\bibitem{Schonfeld:1981kb}
J.~F.~Schonfeld,
Nucl.\ Phys.\ B {\bf 185}, 157 (1981).


\bibitem{Deser:1982vy}
S.~Deser, R.~Jackiw and S.~Templeton,
Phys.\ Rev.\ Lett.\  {\bf 48}, 975 (1982).

\bibitem{Deser:1982wh}
S.~Deser, R.~Jackiw and S.~Templeton,
Annals Phys.\  {\bf 140}, 372 (1982)
[Erratum-ibid.\  {\bf 185}, 406.1988\ APNYA,281,409 (1982)].

\bibitem{Horowitz:1989ng}
G.~T.~Horowitz,
Commun.\ Math.\ Phys.\  {\bf 125}, 417 (1989).

\bibitem{Blau:1991bq}
M.~Blau and G.~Thompson,
Annals Phys.\  {\bf 205}, 130 (1991).

\bibitem{Baez:1994zz}
J.~C.~Baez,
{\sl ``Knots and quantum gravity: Progress and prospects,''}
gr-qc/9410018.

\bibitem{Peldan:1994hi}
P.~Peldan,
Class.\ Quant.\ Grav.\  {\bf 11}, 1087 (1994)


\bibitem{Fucito:1997ax}
F.~Fucito, M.~Martellini and M.~Zeni,
Nucl.\ Phys.\ B {\bf 496}, 259 (1997)

\bibitem{Cattaneo:1998eh}
A.~S.~Cattaneo et al,
Commun.\ Math.\ Phys.\  {\bf 197}, 571 (1998).

\bibitem{Fucito:1997sq}
F.~Fucito et al,
Phys.\ Lett.\ B {\bf 404}, 94 (1997).

\bibitem{Accardi:1997bf}
A.~Accardi and A.~Belli,
Mod.\ Phys.\ Lett.\ A {\bf 12}, 2353 (1997).

\bibitem{Chan:1995bp}
H.~Chan, J.~Faridani and S.~T.~Tsou,
Phys.\ Rev.\ D {\bf 52}, 6134 (1995)

\bibitem{Chan:1996xr}
H.~Chan, J.~Faridani and S.~Tsou,
Phys.\ Rev.\ D {\bf 53}, 7293 (1996)

\bibitem{Lahiri:1992yz}
A.~Lahiri,
Phys.\ Lett.\ B {\bf 297}, 248 (1992).

\bibitem{Lahiri:1992hz}
A.~Lahiri,
{\sl ``Generating vector boson masses,''}
hep-th/9301060.


\bibitem{Cremmer:1974mg}
E.~Cremmer and J.~Scherk,
Nucl.\ Phys.\ B {\bf 72}, 117 (1974).

\bibitem{Aurilia:1981xg}
A.~Aurilia and Y.~Takahashi,
Prog.\ Theor.\ Phys.\  {\bf 66}, 693 (1981).

\bibitem{Govindarajan:1982jp}
T.~R.~Govindarajan,
J.\ Phys.\ G {\bf G8}, L17 (1982).

\bibitem{Allen:1990kc}
T.~J.~Allen, M.~J.~Bowick and A.~Lahiri,
Phys.\ Lett.\ B {\bf 237}, 47 (1990).

\bibitem{Allen:1991gb}
T.~J.~Allen, M.~J.~Bowick and A.~Lahiri,
Mod.\ Phys.\ Lett.\ A {\bf 6}, 559 (1991).

\bibitem{Minahan:1989vc}
J.~A.~Minahan and R.~C.~Warner,
{\sl ``Stuckelberg Revisited,''} Florida U. preprint
UFIFT-HEP-89-15.

\bibitem{Hwang:1997er}
D.~S.~Hwang and C.~Lee,
J.\ Math.\ Phys.\  {\bf 38}, 30 (1997).

\bibitem{Lahiri:1997dm}
A.~Lahiri,
Phys.\ Rev.\ D {\bf 55}, 5045 (1997).

\bibitem{Lahiri:2001uc}
A.~Lahiri,
Phys.\ Rev.\ D {\bf 63}, 105002 (2001).


\bibitem{Teitelboim:1986ya}
C.~Teitelboim,
Phys.\ Lett.\ B {\bf 167}, 63 (1986). 

\bibitem{Henneaux:1997mf}
M.~Henneaux et al,
Phys.\ Lett.\ B {\bf 410}, 195 (1997).

\bibitem{Lahiri:2001di}
A.~Lahiri,
{\sl ``Gauge transformations of the non-Abelian two-form,''}
hep-th/0107104.

\bibitem{Kalb:1974yc}
M.~Kalb and P.~Ramond,
Phys.\ Rev.\ D {\bf 9}, 2273 (1974).


\bibitem{Thierry-Mieg:1983un}
J.~Thierry-Mieg and L.~Baulieu,
Nucl.\ Phys.\ B {\bf 228}, 259 (1983).

\bibitem{Oda:1990tp}
I.~Oda and S.~Yahikozawa,
Phys.\ Lett.\ B {\bf 234}, 69 (1990).

\bibitem{Smailagic:2000hr}
A.~Smailagic and E.~Spallucci,
Phys.\ Lett.\ B {\bf 489}, 435 (2000).

\bibitem{Henneaux:1991my}
M.~Henneaux,
{\it ``On the use of auxiliary fields in classical mechanics and in
field theory,''} 
ULB-TH2-91-04
{\it in Contemporary Mathematics, Proceedings of the 1991
Joint Summer Research Conference on Mathematical Aspects of
Classical Field Theory}.

\bibitem{Lee:1998qu}
C.~Y.~Lee and D.~W.~Lee,
J.\ Phys.\ A {\bf 31}, 7809 (1998).

\bibitem{Harikumar:2001eb}
E.~Harikumar, A.~Lahiri and M.~Sivakumar,
Phys.\ Rev.\ D {\bf 63}, 105020 (2001).

\bibitem{Henneaux:1992ig}
M.~Henneaux and C.~Teitelboim,
``Quantization of gauge systems,''
{\it  Princeton, USA: Univ. Pr. (1992)}.


\bibitem{Barnich:2000zw}
G.~Barnich, F.~Brandt and M.~Henneaux,
Phys.\ Rept.\  {\bf 338}, 439 (2000)




\end{thebibliography}
\end{document}